\documentstyle[12pt,epsfig,rotating]{article}
\def\pge{\pagestyle{empty}} \def\pgn{\pagestyle{plain}}
\def\bsls35{\baselineskip 0.35in}
     \def\d{{\rm d}}
\def\spg{\setcounter{page}} 
\def\bd{
\begin{document}} \def\ed{\end{document}}
\def\bmp{\begin{minipage}} \def\emp{\end{minipage}}
\def\bcc{\begin{center}} \def\ecc{\end{center}}     \def\npg{\newpage}
\def\beq{\begin{equation}} \def\eeq{\end{equation}} \def\hph{\hphantom}
\def\be{\begin{equation}} \def\ee{\end{equation}} \def\r#1{$^{[#1]}$}
\def\n{\noindent} \def\ni{\noindent} \def\pa{\parindent} 
\def\hs{\hskip} \def\vs{\vskip} \def\hf{\hfill} \def\ej{\vfill\eject} 
\def\cl{\centerline} \def\ob{\obeylines}  \def\ls{\leftskip}
\def\underbar#1{$\setbox0=\hbox{#1} \dp0=1.5pt \mathsurround=0pt
   \underline{\box0}$}   \def\ub{\underbar}    \def\ul{\underline} 
\def\f{\left} \def\g{\right} \def\e{{\rm e}} \def\o{\over} 
\def\vf{\varphi} \def\pl{\partial} \def\cov{{\rm cov}} \def\ch{{\rm ch}}
\def\la{\langle} \def\ra{\rangle} \def\EE{e$^+$e$^-$}
\def\bitz{\begin{itemize}} \def\eitz{\end{itemize}}
\def\btbl{\begin{tabular}} \def\etbl{\end{tabular}}
\def\btbb{\begin{tabbing}} \def\etbb{\end{tabbing}}
\def\beqar{\begin{eqnarray}} \def\eeqar{\end{eqnarray}}
\def\\{\hfill\break} \def\dit{\item{-}} \def\i{\item} 
\def\bbb{\begin{thebibliography}{9} \itemsep=-1mm}
\def\ebb{\end{thebibliography}} \def\bb{\bibitem}
\def\bpic{\begin{picture}(260,240)} \def\epic{\end{picture}}
\def\akgt{\noindent{\bf Acknowledgements}}
\def\fgn{\noindent{\bf\large\bf Figure captions}}
\def\fsz{\footnotesize}
\def\ifmath#1{\relax\ifmmode #1\else $#1$\fi}%
\def\rmt{\ifmath{{\mathrm{t}}}} \def\rmcut{\ifmath{{\mathrm{cut}}}}
\newcommand{\QCD}{{\sc qcd}} \newcommand{\NFM}{{\sc nfm}}
\newcommand{\BNL}{{\sc bnl}} \newcommand{\RHIC}{{\sc rhic}} 
\newcommand{\CERN}{{\sc cern}} \newcommand{\LHC}{{\sc lhc}} 
\newcommand{\ALICE}{{\sc alice}} 
\def\pt{{p_{\rmt}}} \def\vf{\varphi} \def\yct{y_{\rmcut}} \def\kt{k_{\rmt}}
\def\levy{L$\acute{\rm e}$vy} \def\renyi{R$\acute{\rm e}$nyi}
\bd
\pge

\hskip8.8cm{\bf Correspondence author:}

\hskip9.8cm LIU, Lianshou

\hskip9.8cm Institute of Particle Physics

\hskip9.8cm Huazhong Normal University

\hskip9.8cm Wuhan, 430079  China

\hskip8.8cm Email: liuls@iopp.ccnu.edu.cn

\hskip8.8cm FAX: 0086 27 87662646
\vskip0.5cm

\cl{\Large A Monte Carlo Study of the Levy Stability and}
\vskip0.3cm

\cl{\Large  Multifractal Spectrum in e$^+$e$^-$ collisions\footnote{
Supported in part by the National Natural Science Foundation of China 
(NSFC) under Projects 19975021 and 90103019.}}

\vskip1.0cm
\cl{\large \  Chen \ \ Gang$^{1,2}$ and Liu Lianshou$^1$ }

\vskip0.3cm
\cl{\small 1 Institute of Particle Physics, Huazhong Normal University,
Wuhan 430079 China}
\cl{2 Department of Physics Jingzhou Teacher's College, Hubei 434100 China}

\date{ }

\vskip1.5cm

\begin{center}
\begin{minipage}{125mm}
\vskip 0.5in
\begin{center}{\Large ABSTRACT}\end{center}
{\hskip0.6cm

The L$\acute{e}$vy stability analysis is carried out for \EE\ collisions
at Z$^0$ mass using Monte Carlo method. The  L$\acute{e}$vy index $ \mu $
is found to be $\mu=1.701\pm 0.043$. 
The self-similar generalized dimensions $D(q)$ and 
multi-fractal spectrum $f(\alpha)$ are presented. The  R$\acute{e}$nyi 
dimension $D(q)$ decreases with increasing $q$. The self-similar 
multi-fractal spectrum is a convex curve with a maximum at $q = 0$, 
$\alpha = 1.169 \pm 0.011$. The right-side part of the spectrum, 
corresponding to negative values of $q$ is obtained through analytical 
continuation.}
\end{minipage}
\end{center}
\vs0.5cm


\vs0.5cm
{\large PACS number: 13.85 Hd}

\hskip1.8cm

\npg \pgn \spg{2}

\section{Introduction} 
The study of fluctuations in particle physics already has a long 
history going back to early cosmic-ray observations~\cite{THB}. Bia\l as and 
Peschanski proposed to study non-statistical fluctuations in 
multi-particle production by the method of factorial moments~\cite{BP}
\beqar   
  F_q(M)&=&{\frac {1}{M}}\sum\limits_{m=1}^{M}{{\langle n_m(n_m-1)
     \cdots (n_m-q+1)\rangle }\over {{\langle n_m \rangle}^q}},
\eeqar
where a region $\Delta$ in 1-, 2- or 3-dimensional phase space is
divided into $M$ cells, $n_m$  is the multiplicity in the $m$th cell,
and $\langle\cdots\rangle$ denotes vertically averaging over the event
sample. 
There are a large variety of experiments showing that $F_q(M)$ possess an 
anomalous scaling behavior~\cite{Wolf}
\beqar   
 F_q(M)  &\propto& (M)^{ \phi(q)}\ \  \quad \quad (M\to \infty) \ \ ,
\eeqar
Note that in \EE\ collisions the fluctuations or power-law scaling exist 
in 3-dimensional phase space and the dynamical fluctuations has been proved 
to be isotropic~\cite{LLLPrd}\cite{CGDatong}\cite{OPAL}. 
This means that the multi-hadron final states in these collisions
are self-similar fractal. 

Since the fractal property of the multi-hadron final states in
\EE\ collisions has been established, the next step is to study
this fractal system in more detail. The aim of the present paper
is to perform the \levy\ stability and multi-fractal spectrum analysis
for the fractal system of  the multi-hadron final states in \EE|
collisions at the Z$^0$ energy --- $\sqrt s = 91.2$ GeV using
Monte Carlo method.

\section{Measurement of intemittency $\phi(q)$}

In our Monte Carlo simulation a total number of 200000 events of
\EE\ collisions at 91.2 GeV are produced by JETSET 7.4 generator.
The primary produced partons considered is a quark-antiquark pair,
moving back to back in the rest frame.
This implies a cylindrical symmetry about the quark direction.
An appropriate frame to study the development of the $q\bar q$
system is therefore the thrust frame~\cite{THR}, the appropriate
variables are ($y, p_t, \varphi$).

 In order to avoid the influence of a non-flat distribution of the
variables $y$, \(p_t\), and \(\varphi\) on the investigation
of the dynamical fluctuations, all variables are transformed into their
corresponding cumulant forms \cite{CUM}.
\beq  
x(y)={{\int_{y_a}^{y} \rho (y)dy}\over {\int_{y_a}^{y_b} \rho (y)dy}} ,
      \qquad  x(p_t)={{\int_{p_{ta}}^{p_t} \rho (p_t)dp_t}\over
         {\int_{p_{ta}}^{p_{tb}} \rho (p_t)dp_t}},
      \qquad  x(\varphi)={{\int_{\varphi_a}^{\varphi} \rho
     (\varphi)d\varphi}\over {\int_{\varphi_a}^{\varphi_b}
        \rho (\varphi)d\varphi}}.
\eeq

The results for the three dimensional FM defined in Eq.(1) are shown in Fig.1.
After omitting the first point~\cite{COP} to eliminate the influence 
of momentum conservation~\cite{YANG}, the three dimensional 
$2 \sim 7$ order ln$F_q$'s versus ln$M$ data are fitted 
very well by a straight line
\beq  
 \ln F_q = \phi(q) \ln M + b ,
\end{equation}
where $\phi(q)$ is the intemittency index. The resulting parameters
and the corresponding $\chi^2/$DF are listed in table I.

\begin{center}
{\bf Table~I} \ \ The fit parameters of 3-D FM for the Jetset7.4
\vskip0.5cm

\footnotesize{   
 \begin{tabular}{|c|c|c|c|}\hline
order $q$ &  $\phi(q)$ &  $b$  &   $\chi^2$/DF  \\ \hline
2& $0.191\pm0.001$ & $-0.304 \pm 0.003$&57/9 \\ \hline
3 &$0.588\pm0.003$ &$-0.914\pm0.009 $&$27/9$ \\ \hline
4 &$1.135\pm0.009$ &$-1.690\pm0.023$ &$20/9$ \\ \hline
5 &$1.784\pm0.024$ &$-2.554\pm0.061$ &$7/9$\\  \hline
6 &$2.435\pm0.091$ &$-3.289\pm0.237$& $4/8$\\  \hline
7 &$3.223\pm 0.197$ &$-4.221\pm 0.519$&$3/4$\\ \hline

 \end{tabular}}
\end{center}
\normalsize

 The results show power-law scaling when phase space is partitioned
isotropically, indicating the existence of self-similar
(isotropic) dynamical fluctuations. Thus 
the multi-hadron final states in \EE\ collisions at $\sqrt s=91.2$ GeV
is a self-similar fractal.
\vs0.4cm 

\begin{center}
\begin{picture}(250,200)
\put(+5,-35)
{
{\epsfig{file=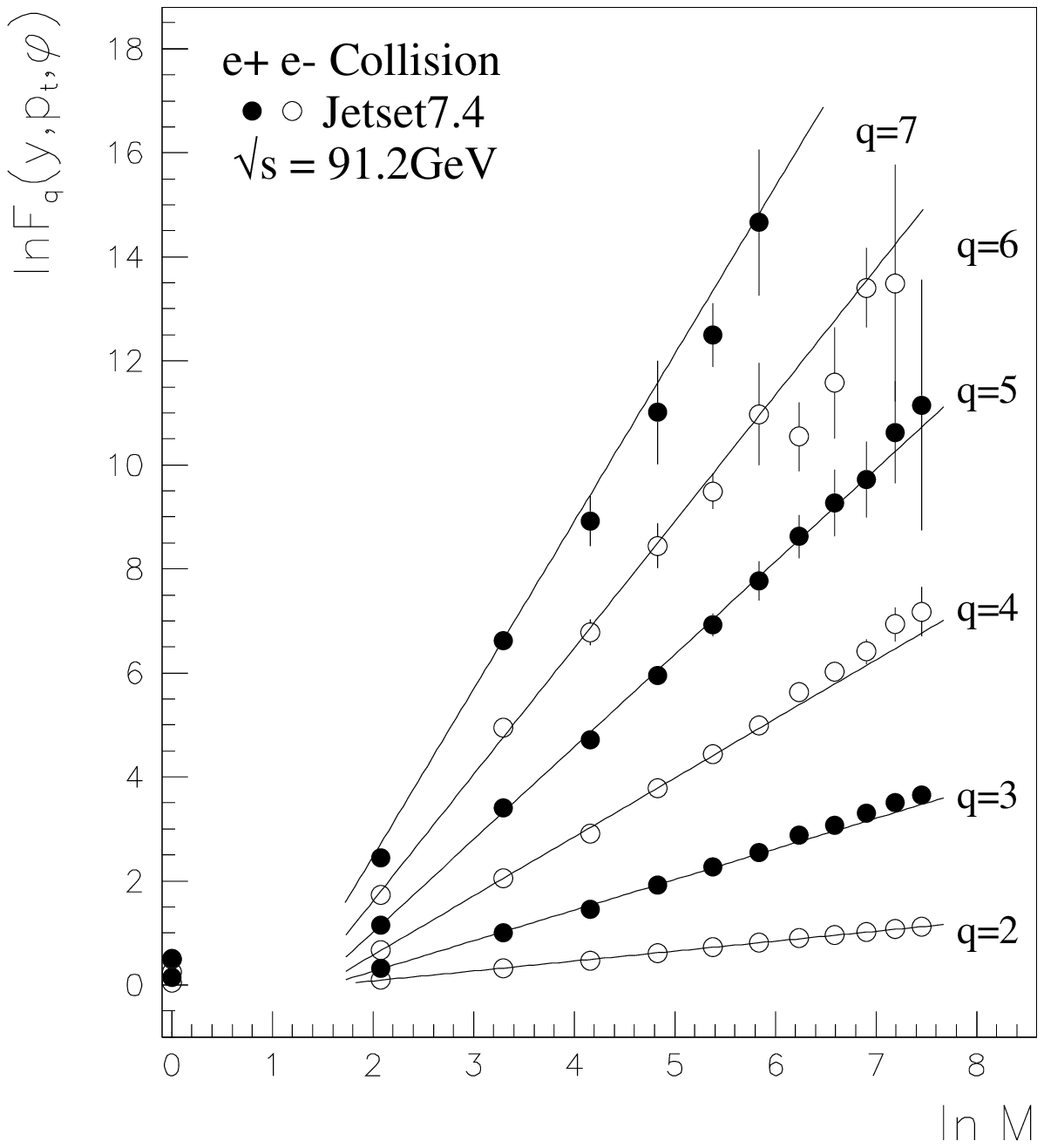,width=230pt,height=258pt}}
}
\end{picture}
\end{center}

\cl{\small Fig.1 \  The three-dimensional $2 \sim 7$ order
 factorial moments as functions of the partition number $M$}                                             
\vskip0.3cm
\section{L$\acute{e}$vy index}

As a self-similar fractal system, the multi-particle final states in high energy
collisions can be charactered by an important parameter --- the L$\acute{e}$vy
 stability index $\mu$. This parameter tells us the behavior of elementary 
fluctuations at the tail. It is also known as the degree of multi-fractality, 
$\mu$ = 0 for mono-fractals, $\mu < 1$ corresponds to the so-called "calm" 
singularities, while $\mu > 1$ correspond to the "wild" singularities.

The intermittency index $\phi (q)$ is related to the anomalous 
dimension $d_q$~\cite{Wolf} as 
\beqar   
 \phi (q) = (q - 1 ) d_q,
\eeqar
and the later is expressible by the L$\acute{e}$vy index $\mu$~\cite{Wolf} as
\beqar   
 d_q = \frac{C_1}{\mu - 1} \frac{q^{\mu} -q}{q - 1} \ \ \ (C_1 = {\rm const}),
\eeqar
or in terms of the intermittency index
\beqar   
 \phi (q) = \phi (2) \ {\frac {q^{\mu } - q}{2^{\mu} - 2}}.
\eeqar

\begin{center}
\begin{picture}(250,320)
\put(-25,40)
{
{\epsfig{file=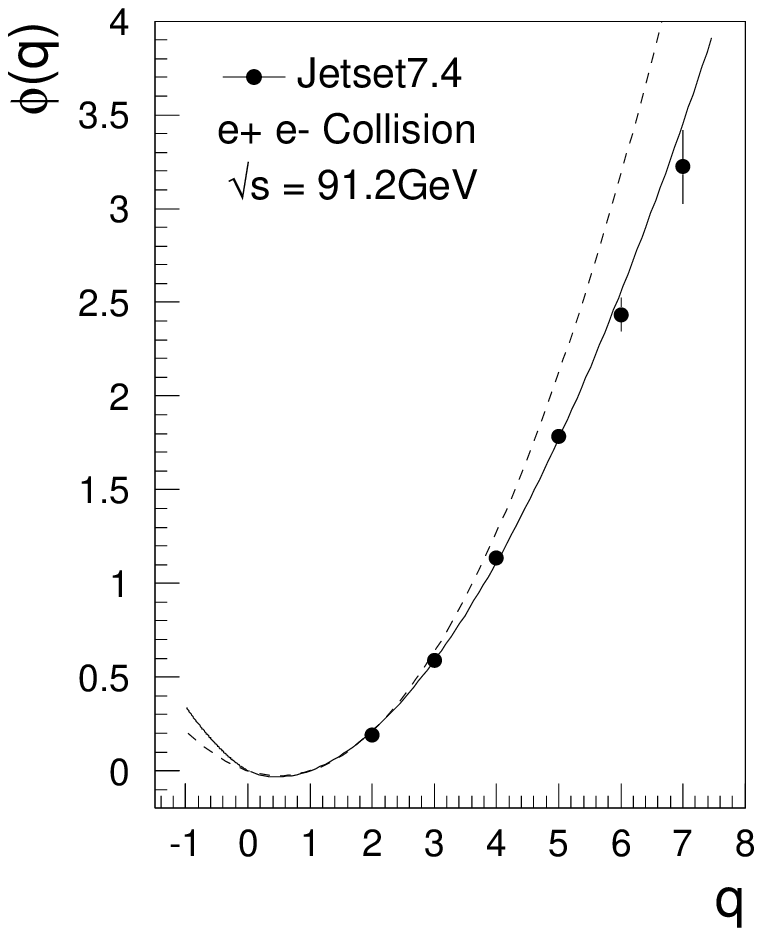,width=320pt,height=328pt}}
}
\end{picture}
\end{center}

\vskip-2.2cm
\cl{{\small Fig.2 \ The intermittency index $\phi _q $
as a function of order $q$.}}  
\cl{{\small Solid curve is from Eq.(9), dashed curve is from Eq.(8).}}
\vskip0.5cm

According to the \levy\ stability theory, the L$\acute{e}$vy index $\mu$ is  
confined to $0 \leq \mu \leq 2$, and the Central Limiting limit corresponds
to $\mu\to 2$. In this limit Eq.(7) passes to
\beqar   
  \phi (q) = \phi (2) \ \frac{q(q-1)}{2}.
\eeqar
Note that Eq.(8) is valid for any real value of $q$, including $q\leq 0$,
while Eq.(7) in the present form is inapplicable to $q\leq 0$. An analytical
continuation of Eq.(7), which is valid for $q\leq 0$ and will, 
in the limit $\mu\to 2$, pass to Eq.(8) in the whole region of $q$ is
\beqar   
 \phi (q) = \phi (2) \ {\frac {({q^2})^{\mu/2} - q}{2^{\mu} - 2}}.
\eeqar 

The distribution of $\phi (q)$ versus $q$ is shown in Table.I and Fig 2. 
We fit it to Eq.(7), shown as solid curve in the figure. This curve has been
extrapolated to $q < 1$ using Eq.(9). Also shown in the figure is the result
from Eq.(8) (dashed curve), which corresponds to the Central Limiting case.
Note that the $q < 1$ part is needed for getting the 
multi-fractal spectrum to the right of its maximum, cf. Fig.3.

The L$\acute{e}$vy index $\mu$ for the hadronic system 
in $e^+ e^-$ collision at $\sqrt s = 91.2$ GeV obtained from the fit is
\beqar 
\mu = 1.701 \pm 0.043. 
\eeqar
This value of $\mu$ lies in the region [0, 2] in consistent
with \levy\ stability.
Since $\mu$ is greater than unity, the multi-fractal corresponds to "wild" 
singularities.

\section{Multi-fractal spectrum and  R$\acute{e}$nyi dimensions} 

The multi-fractal method is a widely used tool in many branches of physics and 
science in general~\cite{YBG}. The R$\acute{e}$nyi
dimension $D_q \ (\equiv 1 - d_q)$ and multi-fractal spectrum $f(\alpha)$ 
are often used to study the multi-fractal. The $D(q)$ (also denoted as
$D_q$) depends on order $q$, being a decreasing function of $q$ for 
multi-fractals in general. The R$\acute{e}$nyi dimension is sometimes called  
as generalized dimension. The multi-fractal spectral function $f(\alpha)$ 
and the R$\acute{e}$nyi dimension (generalized dimension) $D(q)$ for hadronic 
system are related to intermittency index $\phi (q)$ as~\cite{GV}
\beqar   
\tau(q) = q - 1 - \phi (q), \quad  \quad  \alpha(q) = \frac{d\tau (q)}{dq},
\nonumber\\
f(\alpha) = q\alpha - \tau (q),\ \quad  \quad D(q) = \frac{\tau (q)}{(q - 1)}. 
\eeqar
Where $D(q)|_{q=0}  = D_0 = D_F $ is the fractal dimension, 
$ D(q)|_{q=1} = D_I$ is the information dimension and 
$ D(q)|_{q=2} = D_2 = \alpha(2) - f(\alpha(2)) = \nu$ is 
the correlation dimension.

The continuous order index $\phi (q)$ are calculated by Eq.(9).   
Then the R$\acute{e}$nyi dimension $D(q)$ and multi-fractal spectrum 
$f(\alpha)$ are calculated using Eq.(11) with $q$ = -1 to 6.8, step 0.2.   
The $D(q)$ versus $q$ 
and the $f(\alpha)$ versus $\alpha$ are shown in fig.3(a) and (b).

\begin{center}
\begin{picture}(250,280)
\put(-87,0)
{
{\epsfig{file=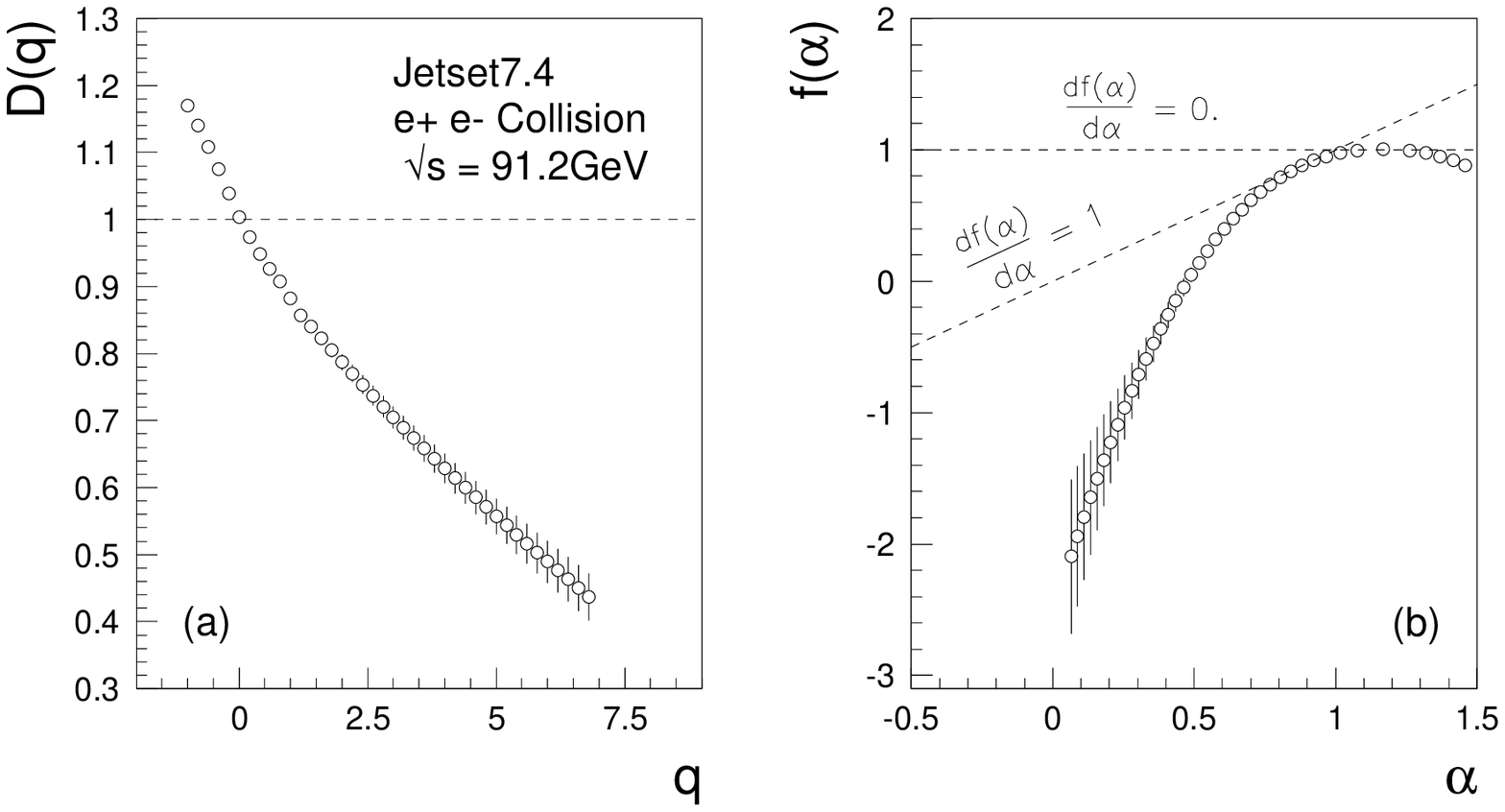,width=450pt,height=288pt}}
}
\end{picture}
\end{center}

\vskip-.5cm
\cl{{\small Fig.3 (a) The  R$\acute{e}$nyi dimension $D(q)$ as a function of order $q$
}}
\cl{{\small \hskip 0.6cm (b)  The multi-fractal spectrum f($\alpha$) as a function of $\alpha$}}

\vskip.3cm
It can be seen from Fig.3 that, the  R$\acute{e}$nyi dimension $D(q)$ 
decreases from 1.17 to 0.437 as the order $q$ increases from $-1$ to 6.8 which 
means that there is a multi-fractal behavior in multi-production at e$^+$e$^-$ 
collisions. The multi-fractal spectrum $f(\alpha)$ is lying
between $-2.10$ and 1 in the above-mentioned $\alpha$ region, 
having a maximum at $q =0$,
$\alpha$ = 1.169$\pm 0.011$. The corresponding fractal dimension is
$D_F = f(\alpha(0)) = D(0) = 1.003 \pm 0.001$. 

The tangents to the curve $f(\alpha) = \alpha$ with slopes 
$\d f(\alpha)/\d\alpha = 1$ and 2 determine the points 
$\alpha(1)=0.882\pm 0.006$ and $\alpha(2)=0.701\pm 0.006$, indicating that
the information dimension of the system is $D_I = D(1) = 0.882 \pm 0.006$,  
and the correlation dimension is
 $ \nu$ = 2$\alpha(2) - f(\alpha(2)) = D(2) = 0.787\pm 0.012$. 

The \renyi\ dimensions $D(q)$ could also be directly calculated using Eq.(11).
The resulting values are listed in Table II together with those obtained
from the Multi-fractal spectrum as described above. It can be seen that 
the consistency is good.
\vskip.2cm
  
\cl{ \bf Table II The \renyi\ dimensions obtained from two different methods} 
\vskip.1cm

\def\btbl{\begin{tabular}} \def\etbl{\end{tabular}}
\bcc\btbl{|c|c|c|c|}\hline
method &$D_F(D(0)) $&$D_I(D(1)) $&$\nu _2(D(2))$ \\ \hline
Equation (11)  
&$1.000\pm0.000$&$0.881 \pm 0.027$&$0.788 \pm 0.003 $  \\ \hline
Multi-fracltal spectrum 
& $1.003\pm 0.001$&$0.882 \pm 0.006$&$0.787 \pm 0.012$  \\  \hline
\etbl \ecc
 \vskip.2cm
  
\section{Conclusions}

A self-similar multi-fractal analysis using MC method in 
\EE\ collisions at $\sqrt s$ = 91.2 GeV 
is performed basing on the measurement of  factorial moments.
The \levy\ stability analysis is carried out and the L$\acute{e}$vy index in 
found to be $\mu = 1.701 \pm 0.043$ consistent with \levy\ stability. 
The multi-fractal spectrum 
$f(\alpha)$ are presented which is a convex curve with a maximum at 
$q = 0.$, $\alpha = 1.169 \pm 0.011$.
The R$\acute{e}$nyi dimensions $D(q)$ are obtained both from their relation
with the intermittency indies $\phi(q)$ and from the multi-fractal spectrum.
The results are in good consistence.

\newpage

\def\Journal#1#2#3#4{{#1} {\bf #2} (#4) #3}
\def\NCA{\em Nuovo Cimento} \def\NIM{\em Nucl. Instrum. Methods}
\def\NIMA{{\em Nucl. Instrum. Methods}| {\bf A}}
\def\NPB{{\em Nucl. Phys.} {\bf B}}
\def\PLB{{\em Phys. Lett.} {\bf B}} \def\PRL{\em Phys. Rev. Lett.}
\def\PRD{{\em Phys. Rev.} {\bf D}} \def\ZPC{{\em Z. Phys.} {\bf C}}

\end{document}